\documentclass{article}
\usepackage{epsfig}
\usepackage{a4wide}

\begin{document}
\title{\vspace*{-1.8cm}Does mesoscopic disorder imply microscopic chaos?}
\author{Peter Grassberger and Thomas Schreiber \\[0.2cm]
Physics Department, University of Wuppertal, Germany}
\maketitle

We argue that Gaspard and coworkers~\cite{G} do not give evidence
for microscopic chaos in the sense in which they use the term. The
effectively infinite number of molecules in a fluid can generate the same
macroscopic disorder without any intrinsic instability. But we argue also
that the notion of chaos in infinitely extended systems needs clarification:
In a wider sense, even some systems without local instabilities can be
considered chaotic. 
~\\

In a beautiful recent experiment~\cite{G}, Gaspard and coworkers verified that
the position of a Brownian particle behaves like a Wiener process with positive
resolution dependent entropy~\cite{GW}.  More surprisingly and
dramatically~\cite{views}, they claim that this observation gives a first proof
of ``microscopic chaos", a term they illustrate by examples of finite
dimensional dynamical systems which are intrinsically unstable.  While the
recent literature finds such chaos on a molecular level quite plausible, we
argue that the observed macroscopic disorder cannot be taken as direct
evidence. In fact, Brownian motion can be derived for systems which would
usually be called non-chaotic, think of a tracer particle in a non-interacting
ideal gas. All that is needed for diffusion is {\it molecular chaos} in the
sense of Boltzmann, i.e. the absence of observable correlations in the motion
of single molecules.

Part of the confusion is due to the lack of a unique definition of
``microscopic chaos" for systems with infinitely many degrees of freedom.  The
authors of~\cite{G} introduce the term by extrapolating finite dimensional
dynamical systems for which chaos is a well defined concept: Initially
close states on average separate exponentially when time goes to infinity.
The rate of separation, the Lyapunov exponent, is independent of the particular
method to measure ``closeness''.  However, the notions of
diffusion and Brownian motion involve by necessity infinitely many degrees of
freedom. In this thermodynamic limit, Lyapunov exponents are no
longer independent of the metric. As a consequence, the large system limit of
a finite non-chaotic system will remain non-chaotic with one particular metric
and become chaotic with another. 

Let us illustrate this astonishing fact with an example introduced by
Wolfram~\cite{wolfram2} in the context of cellular automata. Consider two
states ${\bf x} = (\ldots x_{-2},x_{-1},x_0,x_1,x_2 \ldots)$ and ${\bf y} =
(\ldots y_{-2},y_{-1},y_0,y_1,y_2 \ldots)$ of a one-dimensional bi-infinite
lattice system. If the distance between ${\bf x}$ and ${\bf y}$ is defined by
$d_{\rm max}({\bf x},{\bf y}) = \max_i |x_i-y_i|$, it can grow exponentially
only if the local differences do. This is what one usually means by ``chaos'',
and this is what the authors of~\cite{G} had meant by ``microscopic chaos''.
This mechanism is absent in the thermodynamic limit of finite non-chaotic
systems.  Therefore, some authors~\cite{some} would also call the limit
non-chaotic.  However, the metric $d_{\rm exp}({\bf x},{\bf y}) = \sum_i
|x_i-y_i| \, e^{-|i|}$ can also show exponential divergence if an initially far
away perturbation just moves towards the origin without
growing~\cite{wolfram2}. Arguably, when observing a localised tracer like
in~\cite{G}, the latter choice of metric seems more appropriate.

In finite dimensional dynamical systems, chaos arises due to the de-focusing
microscopic dynamics. The positive entropy is generated by the initially
insignificant digits of the initial condition which are braught forth by the
dynamics.  In the thermodynamic limit, also a completely different mechanism
exists: Perturbations coming from far away regions kick the tracer particle
once and move again away to infinity. The entropy is positive due to
information stored in remote parts of the initial condition. Associated to
this, one can also define suitable Lyapunov exponents~\cite{wolfram2}.

To resolve the confusion, we suggest to follow Sinai~\cite{sinai} and first let
the system size tend to infinity, and only then the observation time. In that
case, a system observed in a particular metric $\mu$ is called $\mu$-chaotic
when we find a positive Lyapunov exponent using this metric.  However, Gaspard
et al. had obviously in mind the type of chaos detectable with the metric
$d_{\rm max}$, and arising from a local instability. For this, they fall short
of giving experimental evidence.

~\\[-2cm]


\begin{thebibliography}{10}
\bibitem{G} 
Gaspard, P., {\sl et al.}
   {\em Nature} {\bf 394}, 865--868 (1998).

\bibitem{views}
D\"urr, D., \& Spohn, H.,
   {\em Nature} {\bf 394}, 831 (1998);
Schewe, F. \& Stein, B. 
   Brownian motion is chaotic.
   {\em Physics News Update} (AIP) Number 389 (Story \#3), September 4, 1998;
Rauner, M. 
   {\em Physikalische Bl\"atter} {\bf 54}, 1001 (1998)

\bibitem{GW}
Gaspard, P., \& Wang, X.-J.
   {\em Phys.\ Rep.} {\bf 235}, 291--345 (1993).

\bibitem{wolfram2} 
Wolfram, S. 
   {\em Physica} {\bf D 10}, 1 (1984).

\bibitem{some}
Crutchfield, J., \& Kaneko, K. 
   {\em Phys.\ Rev.\ Lett.} {\bf 60}, 2715 (1988);
Ershov, S.V., \& Potapov, A.B.
   {\em Phys.\ Lett.} {\bf A 167}, 60 (1992);
Politi, A., Livi, R., Oppo, G.-L., \& Kapral, R. 
   {\em Europhys.\ Lett.} {\bf 22}, 571--576 (1993).

\bibitem{sinai}
   Sinai, Ya.G.
      {\em Int.\ J.\ Bifurcat.\ Chaos} {\bf 6}, 1137 (1996).

\end{thebibliography}
\end{document}